\documentclass[preprint,10pt]{elsarticle}

\makeatletter
\def\ps@pprintTitle{%
\let\@oddhead\@empty
\let\@evenhead\@empty
\def\@oddfoot{\centerline{\thepage}}%
\let\@evenfoot\@oddfoot}
\makeatother

\journal{Physica D}

\usepackage{graphicx,bbm,setspace,amssymb,amsfonts,amsmath,mathtools,mathrsfs,stmaryrd,amsthm,bm,etoolbox,comment,hyperref,url,caption,subcaption,color,enumitem,algorithm,booktabs,microtype,nicefrac,lingmacros,nccmath,tree-dvips,tikz-cd}
\usepackage[noend]{algpseudocode}
\usepackage{algorithm}

\usepackage[colorinlistoftodos]{todonotes}
\usepackage[utf8]{inputenc} 
\usepackage[T1]{fontenc}    

\usepackage[noend]{algpseudocode}

\newtheorem{thm-defn}[theorem]{Theorem/Definition}

\theoremstyle{definition}

\theoremstyle{remark}

\newcommand{\ignore}[1]{}{}

\begin{document}

\begin{frontmatter}

\title{Detecting imbalanced financial markets through time-varying optimization and nonlinear functionals}
   
\author[label1,label2,label3]{Nick James} \ead{nick.james@unimelb.edu.au} \ead{nick.james@bain.com}
\author[label4]{Max Menzies} \ead{max.menzies@alumni.harvard.edu}

\address[label1]{School of Mathematics and Statistics, University of Melbourne, Victoria 3010, Australia}
\address[label2]{Melbourne Centre for Data Science, University of Melbourne, Victoria 3010, Australia}
\address[label3]{Bain \& Company, Sydney 2000, Australia}
\address[label4]{Beijing Institute of Mathematical Sciences and Applications, Beijing 101408, China}

\begin{abstract}

This paper studies the time-varying structure of the equity market with respect to market capitalization. First, we analyze the distribution of the 100 largest companies' market capitalizations over time, in terms of inequality, concentration at the top, and overall discrepancies in the distribution between different times. In the next section, we introduce a mathematical framework of linear and nonlinear functionals of time-varying portfolios. We apply this to study the market capitalization exposure and spread of optimal portfolios chosen by a Sharpe optimization procedure. These methods could be more widely used to study various measures of optimal portfolios and measure different aspects of market exposure while holding portfolios selected by an optimization routine that changes over time.

\end{abstract}


\end{frontmatter}

\section{Introduction}
\label{sec:intro}

The sophistication of the investment business is increasing by the day. The breadth of strategies has brought with it various styles of equity investors, characterized by the way in which they make decisions. Still, various investment styles follow (broadly) similar procedural thinking, and there is relatively high association in the kinds of opportunities to which particular investors allocate capital at any time. Recently, we have experienced equity market regimes where certain stocks are attractive to multiple styles of investors simultaneously. For example, numerous of the largest US companies (by market capitalization) are currently technology companies whose inherent fundamentals (underlying business) and technical (share price) position have lent themselves to various investor types. These include quantitative (momentum-style), growth equity, technology-focused, artificial intelligence (AI) and other thematic factor investments, and more traditional passive index owners. This coincides with a period in which a small number of stocks account for a disproportionate amount of capital in major global markets such as US equities. As the market trends toward this different distributional dynamic, this paper addresses several key questions for investors, including the detection of periods of market imbalance and the properties of well-performing equity portfolios across a range of market capitalization levels.

Ever since the seminal work of Markowitz, Sharpe, and others, emphasizing diversification of portfolios through analysis of asset covariances \cite{Markowitz1952,Sharpe1966}, there has been an extremely rich literature of studying covariance and correlation matrices between assets as a key object of financial market structure. Avenues of research have included principal component analysis of the correlation matrix between equities \cite{Pan2007,Fenn2011,Mnnix2012,Heckens2020}, random matrix theory \cite{Laloux1999,Plerou2002,Gopikrishnan2001}, network analysis \cite{Bonanno2003,Onnela2003,Onnela2004,Utsugi2004,Kim2005,Fiedor2014,Fiedor2014_2} and more.

The literature of portfolio optimization has also grown widely in the same decades as in the above work. There are many approaches to finding a unique optimal portfolio, including statistical mechanics \cite{Zhao2016,Li2021_portfolio,james2021_portfolio}, clustering \cite{Iorio2018,Len2017}, fuzzy sets \cite{Tanaka2000,Ammar2003}, regularization \cite{Fastrich2014,Li2015,Pun2019}, and multiobjective optimization \cite{Lam2021}. NP-hard constraints \cite{Shaw2008,Jin2016} such as portfolio cardinality constraints \cite{Anagnostopoulos2011} may make the selection of a single portfolio a difficult computational problem. In our previous work, we took a contrasting approach and analyzed features and quantiles of random portfolios, rather than attempting to select a portfolio \cite{ james_georg,James2023_cryptoGeorg,James2023_financialcrises, James2024_dislocations,James2024_portfolioEPL}. In this paper, we take a third approach: selecting an optimal portfolio with a standard approach (the Sharpe ratio), and tracking its properties over time.

In comparison to the correlations literature, we believe that mathematical analyses of the structure of the equity with respect to market capitalization (market cap) are less common. \cite{Fort2024} used market capitalization in evolutionary economics to model companies' financial fitness; \cite{Farooq2022,Alshubiri2021} studied the relationship between market cap and the level of investment interest across equities. \cite{Pessa2023} studied the returns of cryptocurrencies related to both market capitalization and age. \cite{Dias2013} and \cite{Wang2014} showed that market cap could be incorporated to improve value-at-risk (VaR) models and momentum trading algorithms, respectively. Market capitalization is of course closely related to price; due to a statistical and mathematical preference for modeling stationary time series, it has been customary to adopt the log return time series of any index or asset class as the primary object of study. We are unaware of other work that studies the multivariate time series of market caps to examine this changing structural aspect of the equity market over time. Further, we believe the literature incorporating market cap data in portfolio optimization is limited, and we wish to add to it from the opposite perspective: taking optimal portfolios and studying their properties with respect to market cap.

Methodologically, our work is principally inspired by a rich literature of applying statistical and physically-inspired models to capture the dynamics of real-world phenomena. In financial markets, these techniques have been applied to a broad range of asset classes including equities \cite{Wilcox2007,james2022_stagflation,james2021_MJW}, foreign exchange \cite{Ausloos2000}, cryptocurrencies \cite{Gbarowski2019,Wtorek2020,James2021_crypto,DrodKwapie2022_crypto,DrodWtorek2022_crypto,DrodWtorek2023_crypto,Drod2023_crypto2}, and debt-related instruments \cite{Driessen2003}. These applied mathematical methods have also been used in a variety of other disciplines including epidemiology \cite{jamescovideu,Manchein2020,Li2021_Matjaz,james2020covidusa,Blasius2020,james2021_TVO,Perc2020,Machado2020,James2020_chaos,Sunahara2023_Matjaz, james2023_covidinfectivity}, environmental sciences \cite{Khan2020,Derwent1995,james2021_hydrogen,Westmoreland2007,james2020_Lp,Grange2018,james2023_hydrogen2,Libiseller2005}, crime \cite{james2022_guns,Perc2013,james2023_terrorist}, the arts \cite{Sigaki2018_art,Perc2020_art}, and other fields \cite{james2021_olympics,Clauset2015,james2021_spectral}. Those interested in recent time series analysis with various societal impacts on the economy should consult \cite{Sigaki2019,Perc_social_physics,Perc2019}.

This paper is structured as follows. In Section \ref{sec:data}, we describe the data analyzed in this paper. In Section \ref{sec:marketcap_structure}, we investigate the structure and decomposition of the market with respect to market capitalization in a time-varying fashion. This can be alternatively interpreted as a means to detect market size imbalances over time. In Section \ref{sec:portfolio_functionals}, we introduce a framework of time-varying linear and nonlinear functionals of optimal portfolios and apply this to investigate the changing total market cap exposure and inequality among optimal portfolios.

\section{Data}
\label{sec:data}

Our data consists of monthly market capitalizations (market cap) and daily stock prices over the last 20 years. Price data ranges from 2003-12-31 to 2024-06-28 inclusive, while market cap data ranges from 2004-01-30 to 2024-06-28, with data reflecting the month-end market cap. Data exists for $n=100$ equities, of which 83 have data ranging for the entire period of analysis.

\section{Market capitalization structural analysis}
\label{sec:marketcap_structure}

In this section, we analyze the distribution of values of market capitalizations over time. We track the distribution of market cap values with time $t$ and directly compare different times in a pairwise manner. Throughout this section, we exclusively study market cap data over monthly time increments.

\subsection{Methodology}

Let $n=100$ be the total number of equities in our data set. The market capitalization data range over a period of $T=246$ months, as detailed in Section \ref{sec:data}. We index the months by $t=1,...,T$. Let $X_i(t)$ be the market capitalization of the $i$th equity at time $t$, for $i=1,...,n$ and $t=1,...,T$. In addition, let $M(t)=\sum_{i=1}^n X_i(t)$ be the total market capitalization of the equities under consideration, interpretable as the total size of the market.

First, we analyze time-varying concentration ratios of the top $k$ equities by market cap. These is defined as follows: for any $t$, order the market caps at that time as $X_{(1)}(t) \geq X_{(2)}(t) \geq ... \geq  X_{(n)}(t)$. Now define the concentration ratio $\text{CR}_k$ as follows:
\begin{align}
\label{eq:concentrationratio}
        \text{CR}_k(t) = \frac{\sum^{k}_{i=1} X_{(i)}(t)}{\sum^n_{i=1} X_i(t)} = \frac{\sum^{k}_{i=1} X_{(i)}(t)}{M(t)},
\end{align}
that is, the proportion of the total market capitalization of our collection of equities concentrated in the top $k$. Any equity that has missing data (for that company does not exist yet) is assigned value 0, effectively dropping it from the concentration ratio calculation. This measure reflects the concentration (at the top) among the distribution of company market capitalizations. We display time-varying (monthly) curves for $k=1,2,3,5,10,20$ in Figure \ref{fig:Concentration_ratios}. We describe and interpret our findings in Section \ref{sec:results_1}.

Next, we investigate the collective imbalance or inequality of market capitalization values, again on a time-varying (monthly) basis. Specifically, at each $t$, we compute the \emph{Gini coefficient} $G(t)$ of the market caps $X_1(t),...,X_n(t).$ This is defined as follows:
\begin{align}
    \label{eq:Gini_def}
    G(t) = \frac{\sum_{i<j} |X_i(t) - X_j(t)|}{n M(t) }.
\end{align}
The numerator is a measure of the absolute deviation between market caps, and when divided by the denominator, $G(t)$ is a normalized quantity in $[0,1]$, with 0 indicating complete equality of all values, and 1 indicating maximal imbalance (all market capitalization concentrated in a single company). $G(t)$ also has a geometric interpretation in terms of the area under the \emph{Lorenz curve}. This relates to the area under an appropriately normalized quantile function of values, where concentration ratios again feature.

Specifically, consider the distribution of values $X_{(1)}(t),...,X_{(n)}(t)$ and normalize them by dividing by the total sum $M(t)$. This yields the previously analyzed concentration ratios. The Lorenz curve is the piecewise linear curve that passes through the points 
\begin{align}
    \left\{ \left( \frac{i}{n}, 1 - \text{CR}_i(t)  \right):    i=0,1,...,n \right\}
\end{align}
This must lie entirely below the line $y=x$ in the unit square $[0,1] \times [0,1]$ and so have area $A$ at most $\frac12$. In fact, the Gini coefficient is related to this area by $G=1-2A$. This reflects the fact that in the case of perfect equality, the Lorenz curve coincides with the line $y=x$, so $A=\frac12$ and $G=0$, while in the case of complete imbalance, $A=0$ and $G=1$. We display the time-varying Gini coefficient $G(t)$ in Figure \ref{fig:Gini}, and describe the results in Section \ref{sec:results_1}. There are two possible approaches to the fact that not all market capitalization exists for all time. One is to compute the Gini coefficient for each time using existing data (dropping missing values for each $t$); the other is to restrict to stocks that have existed throughout the entire window of analysis (dropping those stocks with missingness completely). We performed both calculations and the results were highly similar, indicating the robustness of this approach.

Now we turn our attention to directly comparing the distributions of market capitalization values at different times against each other in a pairwise manner. At each $t$, we recall that we have a distribution of values $X_1(t),...,X_n(t)$. We employ the \emph{Wasserstein metric} between distributions to quantify their difference between any two points of time. Specifically, let $F$ and $G$ be two cumulative distribution functions with quantile functions $F^{-1}$ and $G^{-1}$, respectively. The Wasserstein distance (or more precisely 1-Wasserstein) is defined by
\begin{align}
    \label{eq:Wass1}
    W(F, G)= \int_{0}^1 |F^{-1} - G^{-1}| dx.
\end{align}
This can be generalized to a metric between probability measures $\mu, \nu$ on any metric space, and has a concrete interpretation as the amount of work (in the sense of physics) to move one distribution onto another \cite{DelBarrio}.

With this, we define the difference between any two months $t_1,t_2$ by 
\begin{align}
\label{eq:Wass2}
    D(t_1, t_2)= W(F_{t_1}, F_{t_2}),
\end{align}
where $F_t$ is the cumulative distribution function of normalized values $\frac{X_1(t)}{M(t)},...,\frac{X_n(t)}{M(t)}$. We normalize values in order to remove the effect of market capitalizations simply rising together over time, and instead focus on the distribution of values relative to each other.

If a certain equity does not exist at time $t$, it is implicitly not included in the distribution $F_t$. This is fine for the computation. For times $s$ and $t$ at which all $n$ stocks exist, the metric has a simple form,
\begin{align}
\label{eq:Wass3}
D(t_1, t_2)= \frac{1}{n} \sum_{i=1}^n \left| \frac{X_{(i)}(t_1)}{M(t_1)} - \frac{X_{(i)}(t_2)}{M(t_2)} \right|.
\end{align}
In (\ref{eq:Wass3}), the $i$th largest value $X_{(i)}(t_1)$ at time $t_1$ may not be associated to the same equity as the $i$th largest $X_{(i)}(t_2)$ at time $t_2$. In Figure \ref{fig:D_ST_Clustering}, we display the results of hierarchical clustering on $D(t_1,t_2)$ as $t_1$ and $t_2$ range over all months of the analysis window.

\clearpage

\subsection{Results}
\label{sec:results_1}

\begin{figure*}
    \centering
    \includegraphics[width=1\textwidth]{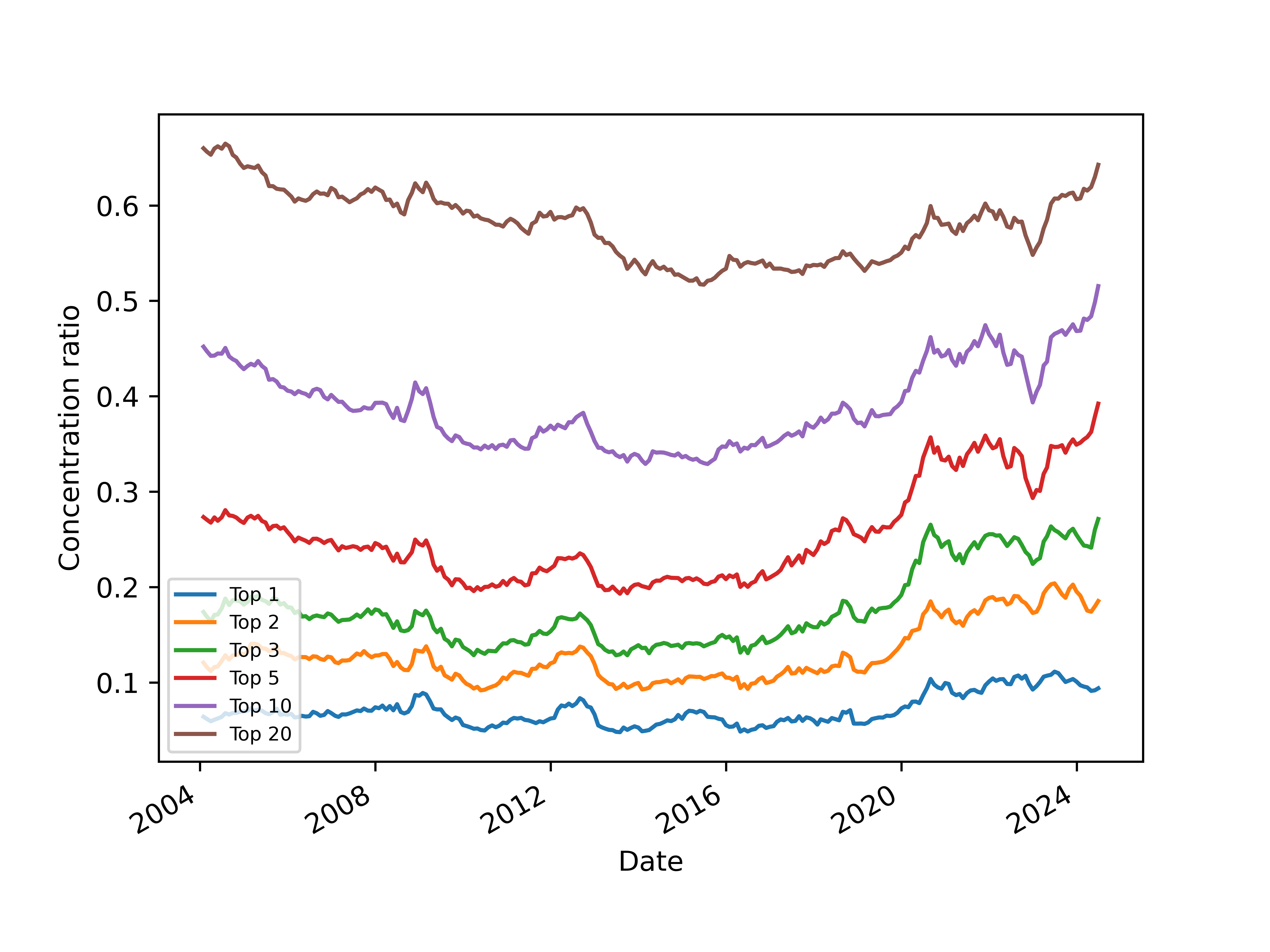}
    \caption{Concentration ratios $\text{CR}_k$ of market capitalizations on a monthly basis for $k=1,2,3,5,10,20$. These curves represent how much of the total market capitalization of the entire market is captured within the top $k$ companies.}
    \label{fig:Concentration_ratios}
\end{figure*}

\begin{figure*}
    \centering
    \includegraphics[width=1\textwidth]{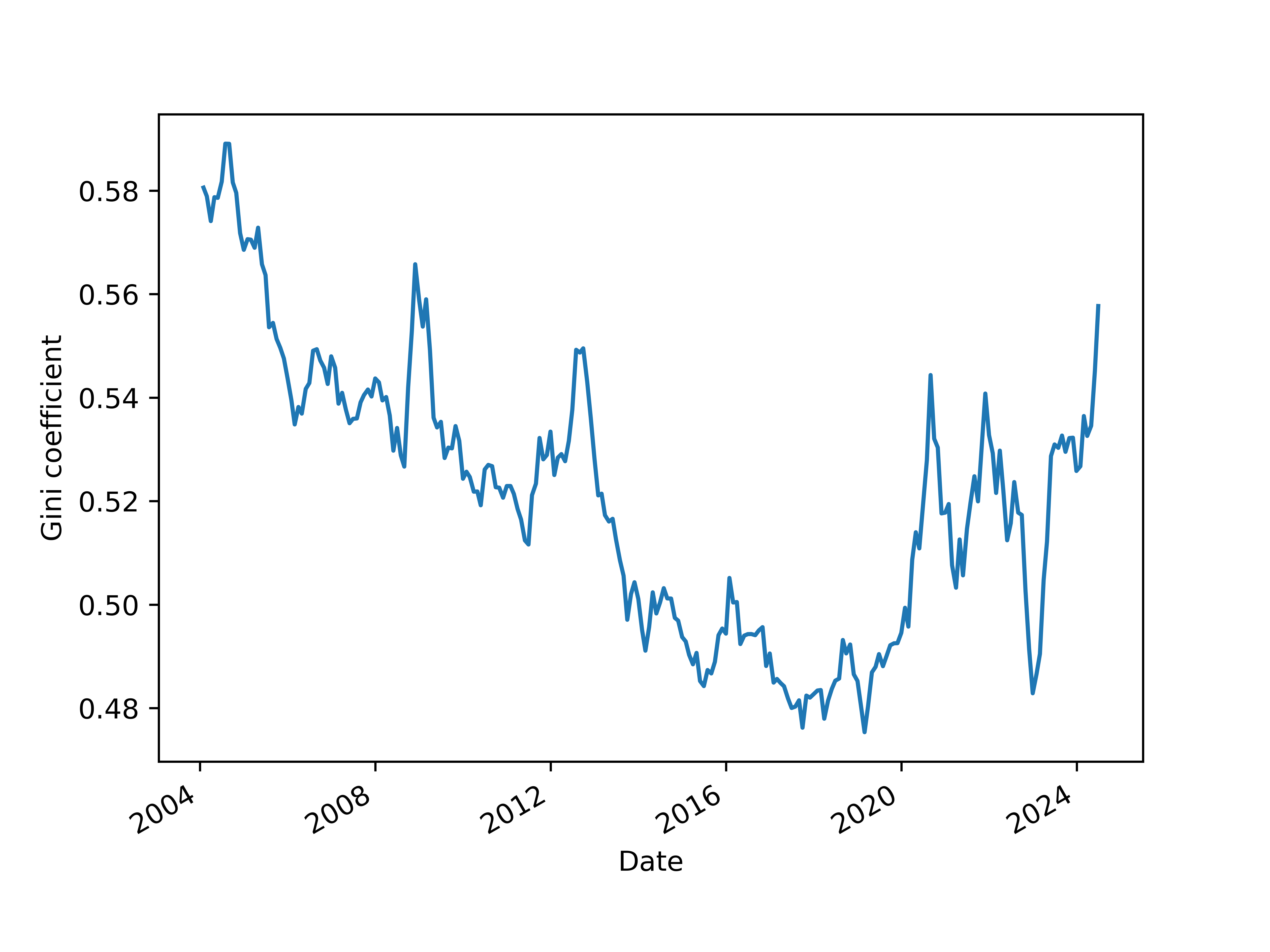}
    \caption{Gini coefficient $G(t)$ of market capitalization data at each month $t$. Lower values indicate greater collective equality among market capitalization values; higher values indicate greater imbalance. We see that inequality among market caps was collectively decreasing until 2020, since which it has exhibited an increase.}
    \label{fig:Gini}
\end{figure*}

\begin{figure*}
    \centering
    \includegraphics[width=1\textwidth]{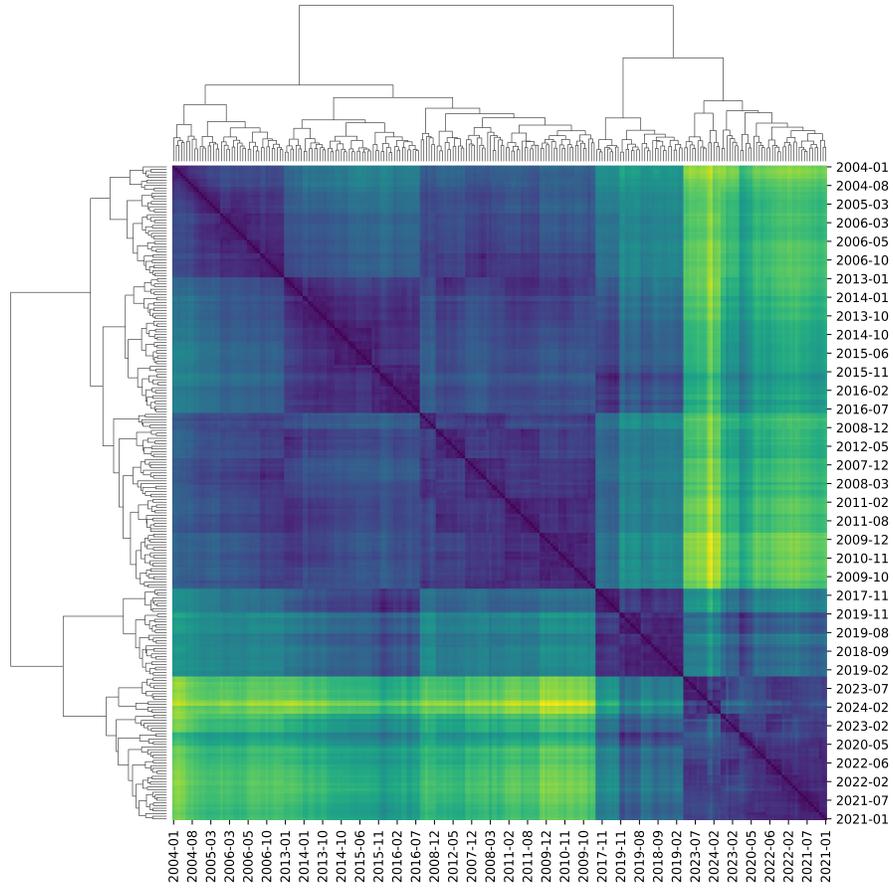}
    \caption{Hierarchical clustering on $W(F_{t_1},F_{t_2})$, defined in (\ref{eq:Wass2}), measuring the discrepancy between distributions of normalized market capitalizations in different months. The greatest differences are observed between pre- and post-2020 periods, mirroring insights from Figures \ref{fig:Concentration_ratios}  and \ref{fig:Gini}. }
    \label{fig:D_ST_Clustering}
\end{figure*}

In Figure \ref{eq:concentrationratio}, we plot the time-varying concentration ratios of the top $k$ companies for  values where $k \in \{ 1, 2, 3, 5, 10, 20 \}$. There are several observations of interest. 
First, all the curves (especially for $k=1,2,3,5$) demonstrate a marked increase around 2020, coinciding with COVID-19 and the associated financial crisis, with values consistently high since then and even an additional uptick at the end of our analysis window. This demonstrates market cap distribution becoming more tightly concentrated in the $\sim 5$ largest companies, with a longer tail of companies who are (relatively) smaller. Indeed, by the end of our analysis window, holding a portfolio of $k=5$ companies contains almost 40\% of the total market concentration (across our 100 stocks). This has numerous implications for investment decisions. On the one hand, investors can attain predominant market exposure relatively cheaply (with far fewer than 100 stocks). On the other hand, this also indicates a degree of risk in some more passive and index-based investment strategies, where such a high percentage of an investor's allocation can be held in just a few companies. Next, the curves give us a visual indication of the cumulative density function of the market's concentration of size. Specifically, the curves in Figure \ref{eq:concentrationratio} are approximately equally spaced, showing an approximately uniform distribution of market capitalization between the values $k=1,2,3,5,10,20$. We can also observe certain changes in how these distributions evolve with time. We can observe there has been a narrowing in the distribution between 2004-2024 for the 5th-20th largest companies.

Figure \ref{fig:Gini} tells a similar story, demonstrating a similar down-then-up pattern as in Figure \ref{eq:concentrationratio}. The trend in the Gini coefficient shows that the market exhibited the greatest inequality at the start and end of the data period under analysis, with a more balanced market cap distribution between 2012 and 2020. Mirroring Figure \ref{eq:concentrationratio}, collective inequality has increased since the start of 2020. It is worth comparing the Gini coefficient with the concentration ratios previously discussed. The Gini coefficient measures overall inequality, so this is an additional (supportive) finding, rather than a restatement of previous results. It is indeed conceivable that this change in the Gini coefficient is primarily driven by the top companies given their disproportionate size (seen in Figure \ref{eq:concentrationratio}).

We now turn to Figure \ref{fig:D_ST_Clustering}, which clusters the distance matrix $D(t_1, t_2)$ between the distribution of normalized company market capitalizations at all points in time. This figure is a novel approach for comparing different periods, where investors can address questions surrounding the similarity of locally stationary segments or market regimes. The dendrogram reveals a striking difference in the distribution function between the dates 2020-2024 and the rest of the analysis window. This supports our findings from Figures \ref{fig:Concentration_ratios} and \ref{fig:Gini}, with a changing dynamic in market capitalizations since 2020. Further, we see four distinct subclusters around periods 2004-2006, 2007-2012, 2013-2016, 2017-2019, representing a strong temporal dependence. While some grouping of adjacent times is to be expected (due to normalized market caps changing continuously with time), there are still some insights here. First, the market concentration dynamics appear to change at a reasonably consistent rate, with all these subclusters containing about four years of data. More surprising, when breaks between the clusters are observed, the subsequent period is not necessarily most similar to adjacent periods. For example, the 2004-2006 period has higher affinity with the post-GFC period than the GFC period itself (which is nearer in time). Overall, the most significant change in market capitalization structure occurred around 2020.

It is worth comparing this primary finding with the much wider literature, summarized in Section \ref{sec:intro}, which studies the structure of correlations between assets over time. For technical reasons, correlation (especially on a time-varying basis) must be computed between stationary time series, which is the primary reason researchers typically study the time series of log returns. Typically, such papers share a common finding: correlations (and other market aspects) change drastically during financial crises. This was observed early by \cite{Drod2000} and in many other works such as \cite{James2021_crypto2,James2024_dislocations}, where the collective strength of correlations increases during crises, for both equity and cryptocurrency markets. In this paper, we study the structure of market capitalizations of different equities; market capitalization is closely related to price, which effectively integrates over daily log returns. Thus, it should be no surprise that the form of our results are a little different, with drastic changes observed since 2020 (for example), rather than just within 2020, as would be observed for correlations.

\clearpage

\section{Time-varying linear and nonlinear functionals of optimal portfolios}
\label{sec:portfolio_functionals}

The previous section studied the structure of the entire market over time with respect to market capitalization. In this section, we turn our attention to portfolios formed from this collection of stocks, and analyze several aspects of optimal portfolios as they change over time. We develop a theory of linear and nonlinear functionals of optimal portfolios, focused on how they relate to market capitalization size and spread. We make use of both daily price data and monthly market cap data in this section.

\subsection{Methodology}

First, we recall the \emph{Sharpe ratio} optimization problem as a means to select optimal portfolio weights. Given a collection of $m$ assets and data over some period, let $R_i$ be the historical returns for the $i$th asset, $\Sigma$ be the historical covariance matrix between the assets, $R_{f}$ the risk-free rate (which we set to 0), and $w_i$ the weights of the portfolio. One maximizes the Sharpe ratio in the following optimization problem:
\begin{align}
\label{eq:Sharpeobjectionfn}
\text{Maximize: } \frac{\sum^{m}_{i=1} w_{i} R_{i} - R_f}{ \sqrt{\boldsymbol{w}^{T} \Sigma \boldsymbol{w}}  }, \\
\text{subject to: } 0 \leq w_{i} \leq b, i = 1,...,m, \label{constrant1} \\
\sum^{m}_{i=1} w_{i} = 1.\label{constrant2}
\end{align}
The Sharpe ratio measures the risk-adjusted return of the portfolio, and the optimization problem above selects a long-only allocation of weights $w_i \geq 0$. The constraint $w_i \leq b$ provides upper bounds on the weights $w_i$ to avoid excessive holdings in individual assets and to comply with asset allocation guidelines. We set $b=0.15$ in all experiments reflecting a 15\% weight cap of any individual asset.

Before we proceed, we note some differences in our two data sets. Price data exist on a daily basis (trading days only) from 2003-12-31 to 2024-06-28. After computing daily returns, returns data exist from 2004-01-02 to 2024-06-28, a period of $S=5157$ days. On the other hand, market capitalization data exists for the same months (on the final trading day of each) from 2004-01-30 to 2024-06-28, a period of $T=246$ months. To disambiguate, we use $s=1,...,S$ to index daily data, while $t=1,...,T$ will index monthly data, coinciding with the notation of Section \ref{sec:marketcap_structure}. To avoid misleading trends in weights starting from 0 for equities that don't exist yet, we restrict this section to the $m=82$ equities that existed throughout our window of analysis for both returns and market cap data.

First, we compute time-varying optimal portfolio weights over a rolling window of $\rho=180$ days. That is, for $s=\rho,...,S$, we consider returns $R_i, i=1,...,m$ over a time interval $[s-\rho + 1, s]$ and compute optimal weights $w_i(s)$. Figure \ref{fig:Heatmap_weight} shows a heat map depicting all the weights $w_i(s)$ over time.

Next, we aim to measure pairwise discrepancy between the weight trajectories of individual assets. Given the consistent constraint that weights must sum to 1, we use the $L^1$ norm between weight trajectories (considered as vectors) defined by
\begin{align}
    \label{eq:L1norm}
    \| w_i - w_j \| = \sum_{s=\rho}^S |w_i(s) - w_j(s)|, i, j=1,...,m,
\end{align}
and perform hierarchical clustering based on this measure, depicted in Figure \ref{fig:weight_clustering}.

We now turn to an analysis not merely of the individual weights but of the optimal portfolios as a whole. The idea is to define functionals, which may be linear or nonlinear, of the entire portfolio at every time. Recall that a functional is a scalar-valued map $f: \mathbb{R}^m \to \mathbb{R}$ on a (high-dimensional) vector space. We wish to apply various functionals to the weight vectors of optimal portfolios over time. That is, we want to compute time-varying $f(t)=f(w_1(t),...,w_m(t))$ to gain inference into broad trends in the changing nature of portfolios.

In what follows, we utilize both market cap and price data and compute both linear and nonlinear functionals that reflect market cap total exposure and spread of optimal portfolios. Our first functional is linear. For any month $t$, we define
\begin{align}
    \label{eq:functional1}
    \nu(t)= w_1(t)X_1(t) + ... + w_m(t) X_m(t).
\end{align}
This calculates the total market exposure at time $t$ using weights computed over the prior six months. We also slightly modify this naive definition. Let $\tau=6$ and again consider the previous $\tau=6$ months prior to and including $t$. Then we define
\begin{align}
\label{eq:functional2}
\bar{X}_i(t) = \frac{1}{\tau} \sum_{u=t - \tau + 1}^t X_i(u); \\
    \bar{\nu}(t)= w_1(t)\bar{X}_1(t) + ... + w_m(t) \bar{X}_m(t).
\end{align}
In (\ref{eq:functional2}), we use the average market cap for each asset over the previous six months, the same period over which the optimal Sharpe portfolio weights are selected. This reflects a more appropriate measure of market cap total exposure of a chosen portfolio over a period. We remark that in the case of equal portfolio weightings $w_i=\frac{1}{m}$ for all $i$, $\nu(t)$ coincides with $\mu(t)=\frac{1}{m} M(t)$, the average market cap of all stocks in consideration, while $\bar{\nu}(t)$ coincides with $\bar{\mu}(t)=\frac{1}{m} \bar{M}(t)$ where we further average over the last 6 months. Thus, we also report the \emph{normalized market exposure} defined by
\begin{align}
\label{eq:functional3}
    f(t)= \frac{\bar{\nu}(t)}{\bar{\mu}(t)},
\end{align}
which gives $\bar{\nu}(t)$ normalized by the exposure to a uniform portfolio. All these functionals ($\mathbb{R}^m \to \mathbb{R}$) are linear in the weights for each month $t$. In Figure \ref{fig:portfolio_weighted_size}, we display the market cap total exposure $\bar{\nu}(t)$ for months $t$ against the market average $\mu(t)$. In Figure \ref{fig:portfolio_normalized_weights}, we display the normalized market exposure $f(t)$.

Next, we compute a nonlinear functional of optimal portfolios, the time-varying Gini coefficient of portfolio market cap exposures. This is defined as the Gini coefficient of the market capitalization values $\bar{X}_i(t)$ (averaged over the last six months) considered as a discrete distribution with weights $w_i(t)$. This can be defined as
\begin{align}
    \label{eq:functionalGini}
    g(t)=\frac{2}{\bar{\nu}(t)} \sum_{i,j=1}^m w_i(t) w_j(t) |\bar{X}_i(t) - \bar{X}_j(t)|.
\end{align}
We recall that $\bar{\nu}(t)$, as defined above, is simply the mean of the market capitalization values considered as a discrete distribution, and this features in the Gini coefficient as a normalizing term. As in Section \ref{sec:marketcap_structure}, this also has an interpretation in terms of the Lorenz curve of the distribution. We display the time-varying nonlinear portfolio functional $g(t)$ in Figure \ref{fig:portfolio_Gini}.

Finally, we adapt the hierarchical clustering of Section \ref{sec:marketcap_structure} to the setting of optimal portfolios. We recall the Wasserstein metric $W$ between each month's cumulative distribution function $F_t$ of normalized market capitalization values $\frac{X_1(t)}{M(t)},...,\frac{X_n(t)}{M(t)}$ in Eq. (\ref{eq:Wass1}) and (\ref{eq:Wass2}). We adopt this with two alterations. For each month $t=\tau,...,T$, let $H_t$ be the cumulative distribution function of normalized market cap values over the prior $\tau=6$ months, $\frac{\bar{X}_1(t)}{\bar{M}(t)},...,\frac{\bar{X}_n(t)}{\bar{M}(t)}$, using the notation of the previous section, considered as a discrete distribution with weights $w_i(t)$. That is, $H_t$ is the cumulative distribution function whose weights (or value probabilities) are drawn from an optimal Sharpe portfolio (calculated over a 6-month window), and whose values are normalized market caps over the same period. Just like in Section \ref{sec:marketcap_structure}, we normalize to remove the effect of market capitalizations simply rising together over time, and instead focus on the (weighted) distributions of market cap values.

We may then define the difference these distributions at any two times (months) $t_1,t_2$ by 
\begin{align}
\label{eq:Wass_portfolio}
    D(t_1, t_2)= W(H_{t_1}, H_{t_2}),
\end{align}
and display hierarchical clustering in Figure \ref{fig:dst_portfolio_weights}.

\clearpage

\subsection{Results}

\begin{figure*}
    \centering
    \includegraphics[width=1\textwidth]{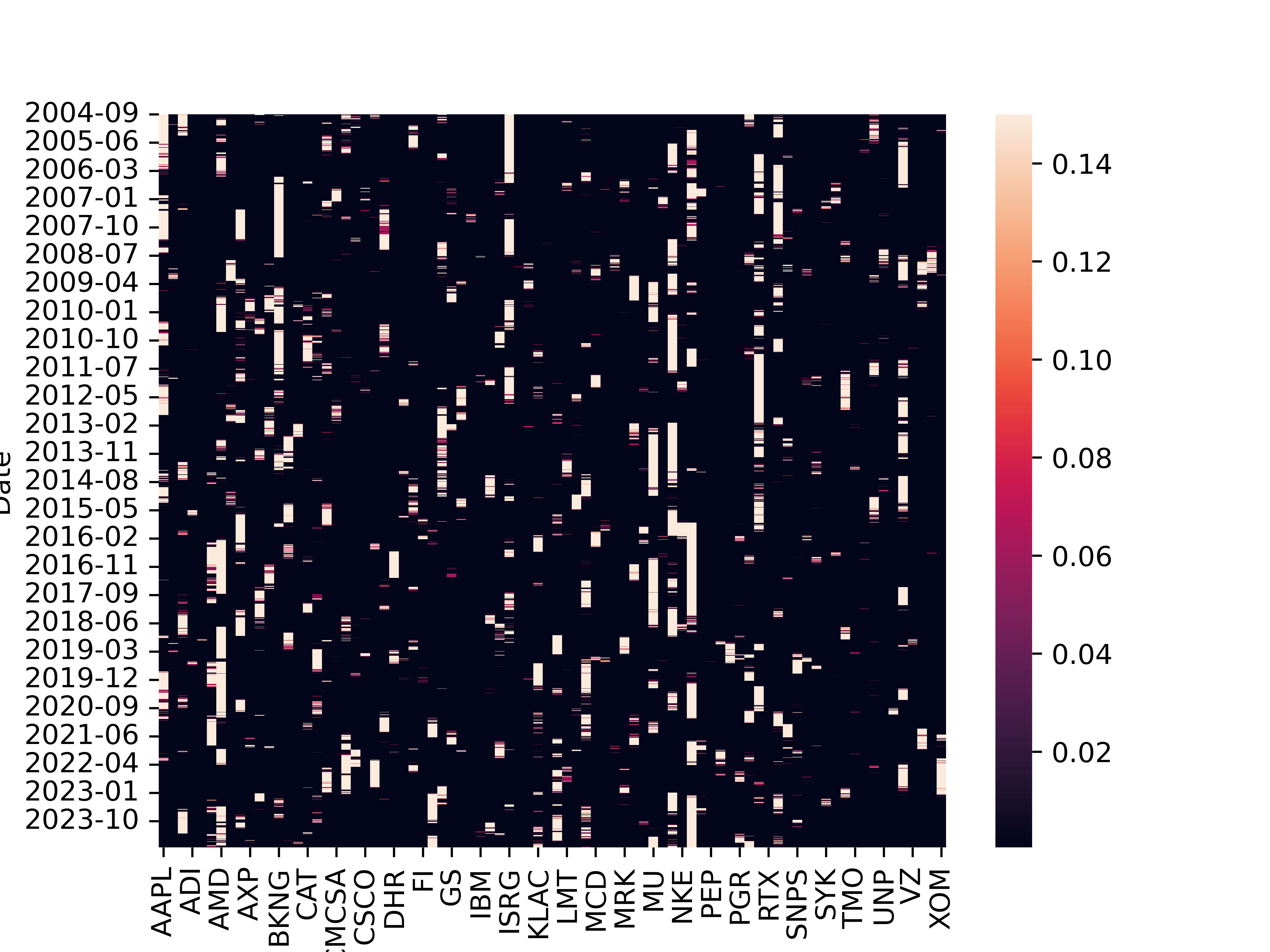}
    \caption{Heatmap of time-varying optimal weights $w_i(s)$ as determined by the Sharpe ratio maximization procedure over a rolling window of 180 days. The upper bound of any individual securities weight within the portfolio is 15\%.}
    \label{fig:Heatmap_weight}
\end{figure*}

\begin{figure}
    \centering
    \includegraphics[width=\textwidth]{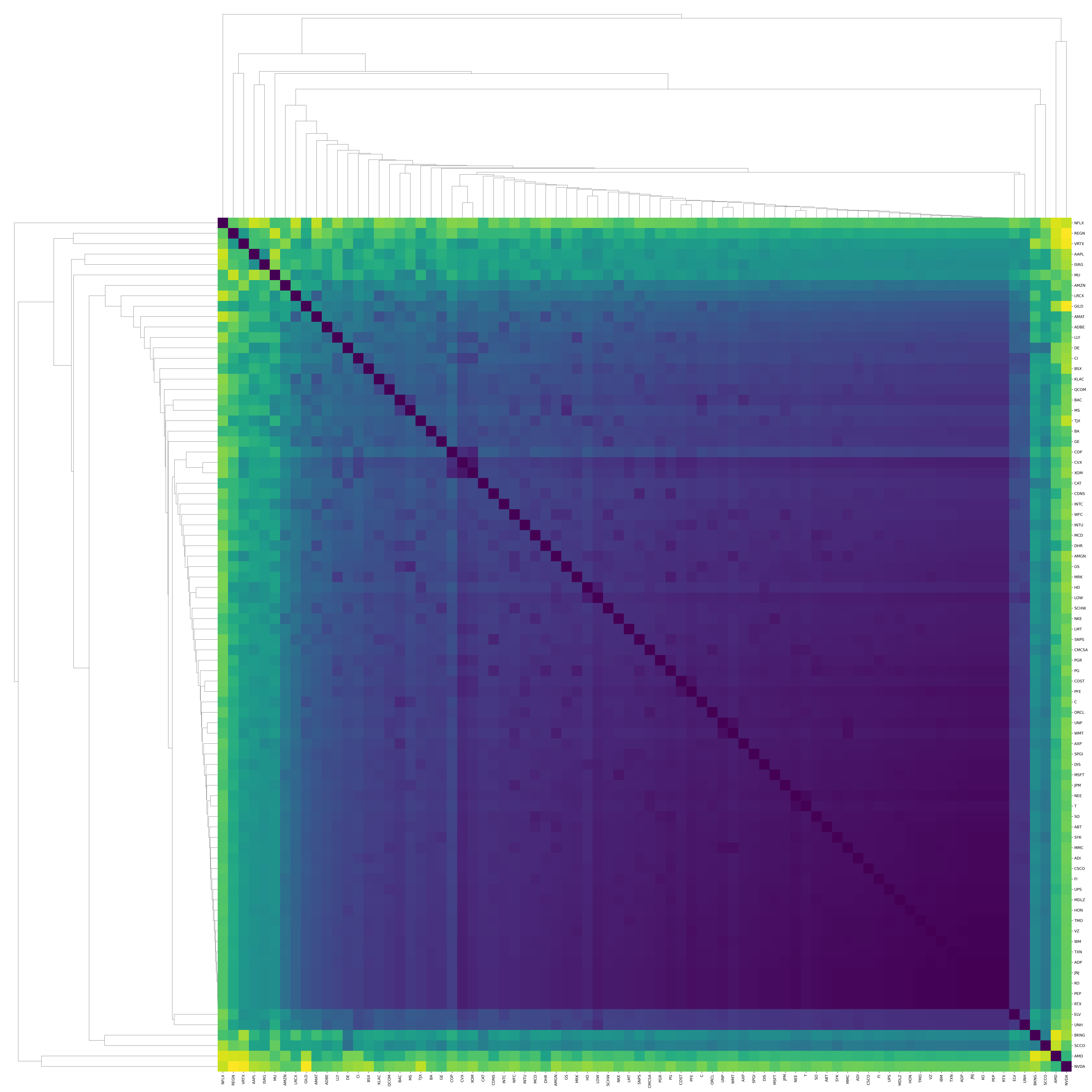}
\caption{Hierarchical clustering based on the $L^1$ norm between weight trajectories (\ref{eq:L1norm}). The outlier on the far left is Netflix. Next to that (on the left side) is a cluster of two, Regeneron and Vertex Pharmaceuticals. Next to that (still on the left side) is another cluster of two, Apple and Intuitive Surgical and other technology companies. The two paired on the far right are Nvidia and AMD.}
\label{fig:weight_clustering}
\end{figure}

\begin{figure}
    \centering
    \includegraphics[width=\textwidth]{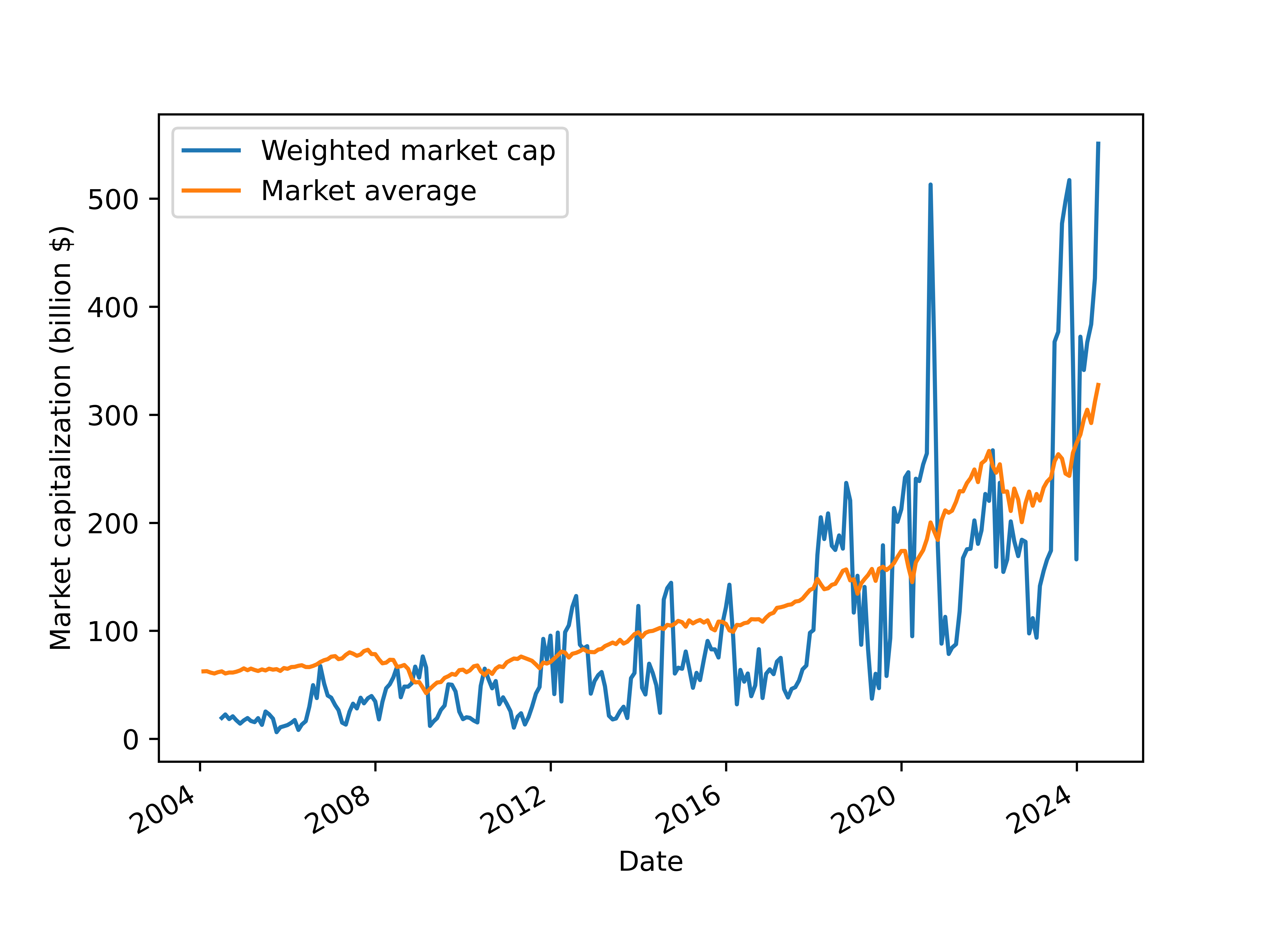}
\caption{The orange curve is the average market capitalization over time $\bar{\mu}(t)$; the blue curve is the weighted market cap of the optimal Sharpe portfolio $\bar{\nu}(t)$ defined in (\ref{eq:functional2}), that is, the total exposure of the portfolio to market cap size. We see the optimal portfolio is generally under-exposed to market capitalization for most of the period of analysis, which switches closer to the present day.}
\label{fig:portfolio_weighted_size}
\end{figure}

\begin{figure}
    \centering
    \includegraphics[width=\textwidth]{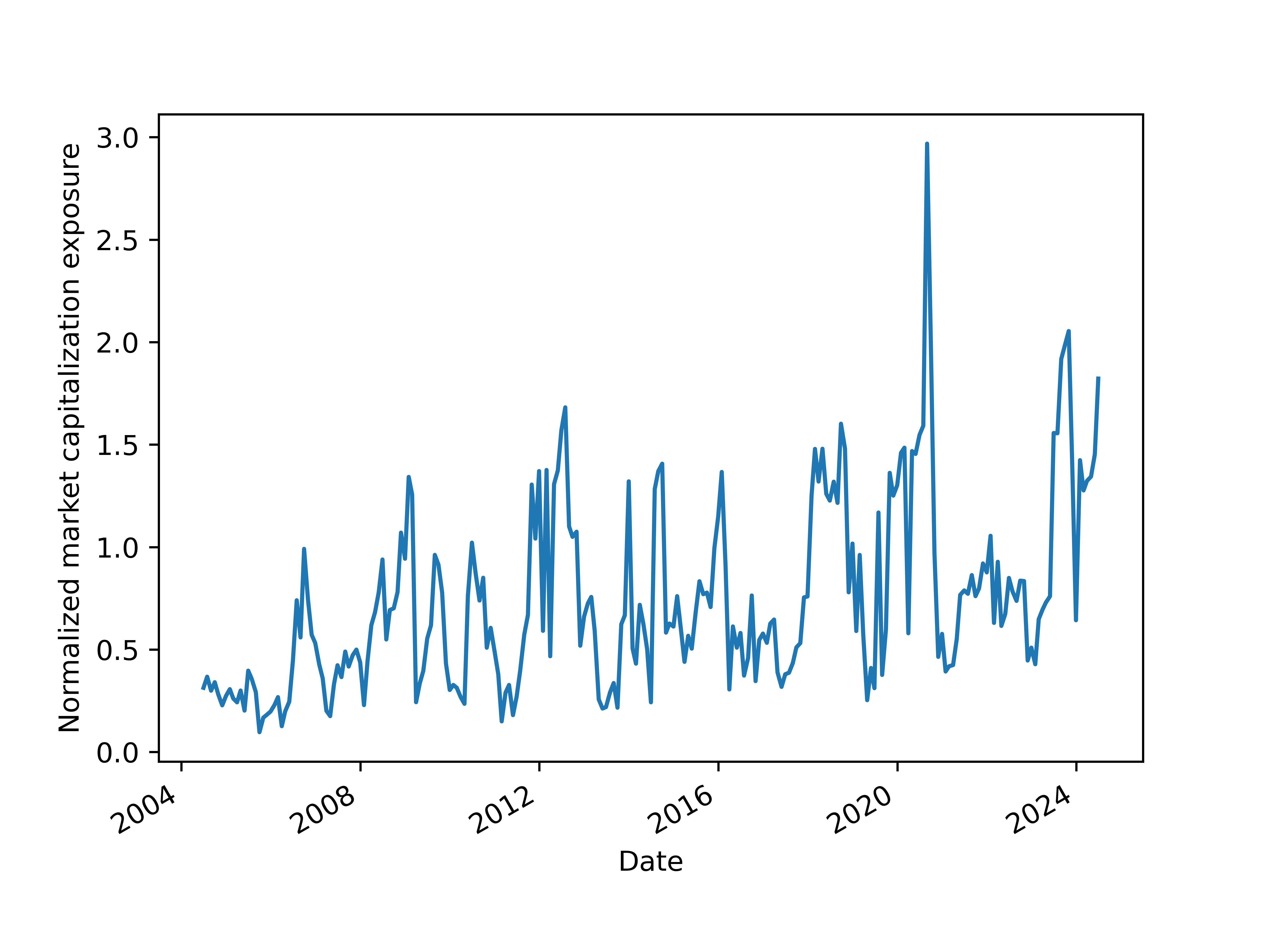}
\caption{This curve shows the weighted market capitalization of the optimal Sharpe portfolio normalized by the average market cap at that time. This coincides with the quotient of the blue and orange curves in Figure \ref{fig:portfolio_weighted_size}, and is defined as $f(t)$ in (\ref{eq:functional3}). We may more closely observe periods in which the optimal portfolio is under- and over-exposed to market size.}
\label{fig:portfolio_normalized_weights}
\end{figure}

\begin{figure}
    \centering
    \includegraphics[width=\textwidth]{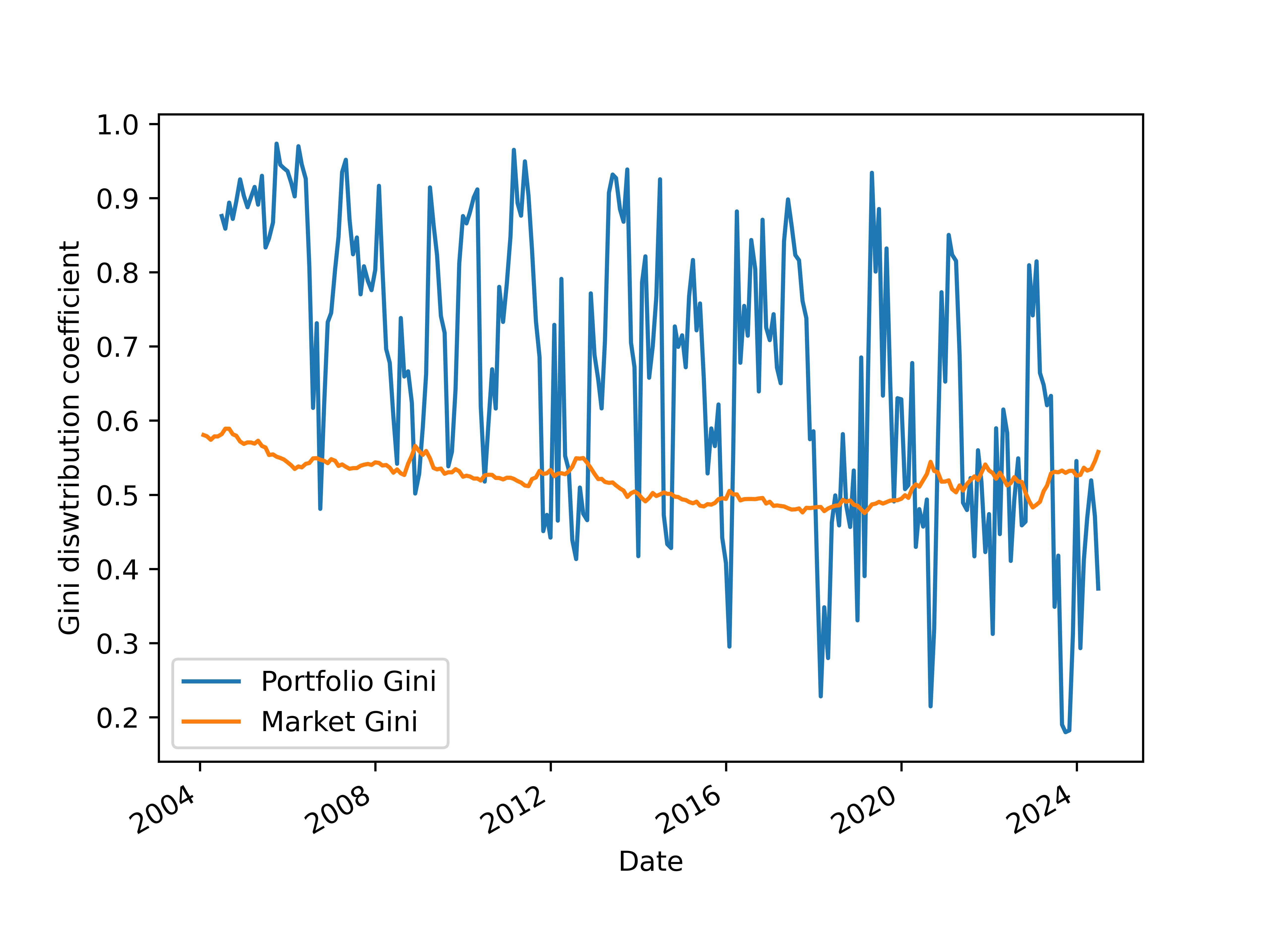}
\caption{The blue curve shows the Gini coefficient $g(t)$ of the distribution of market capitalization values weighted according to the Sharpe portfolio distribution weights, defined in (\ref{eq:functionalGini}. It is a measure of the inequality in the Sharpe portfolio considered as a distribution. The orange curve shows the market Gini $G(t)$ defined in (\ref{eq:Gini_def}), the same as Figure \ref{fig:Gini}). Amid considerable variability, we see a decreasing trend in the optimal portfolio's Gini coefficient with time.}
\label{fig:portfolio_Gini}
\end{figure}

\begin{figure}
    \centering
    \includegraphics[width=\textwidth]{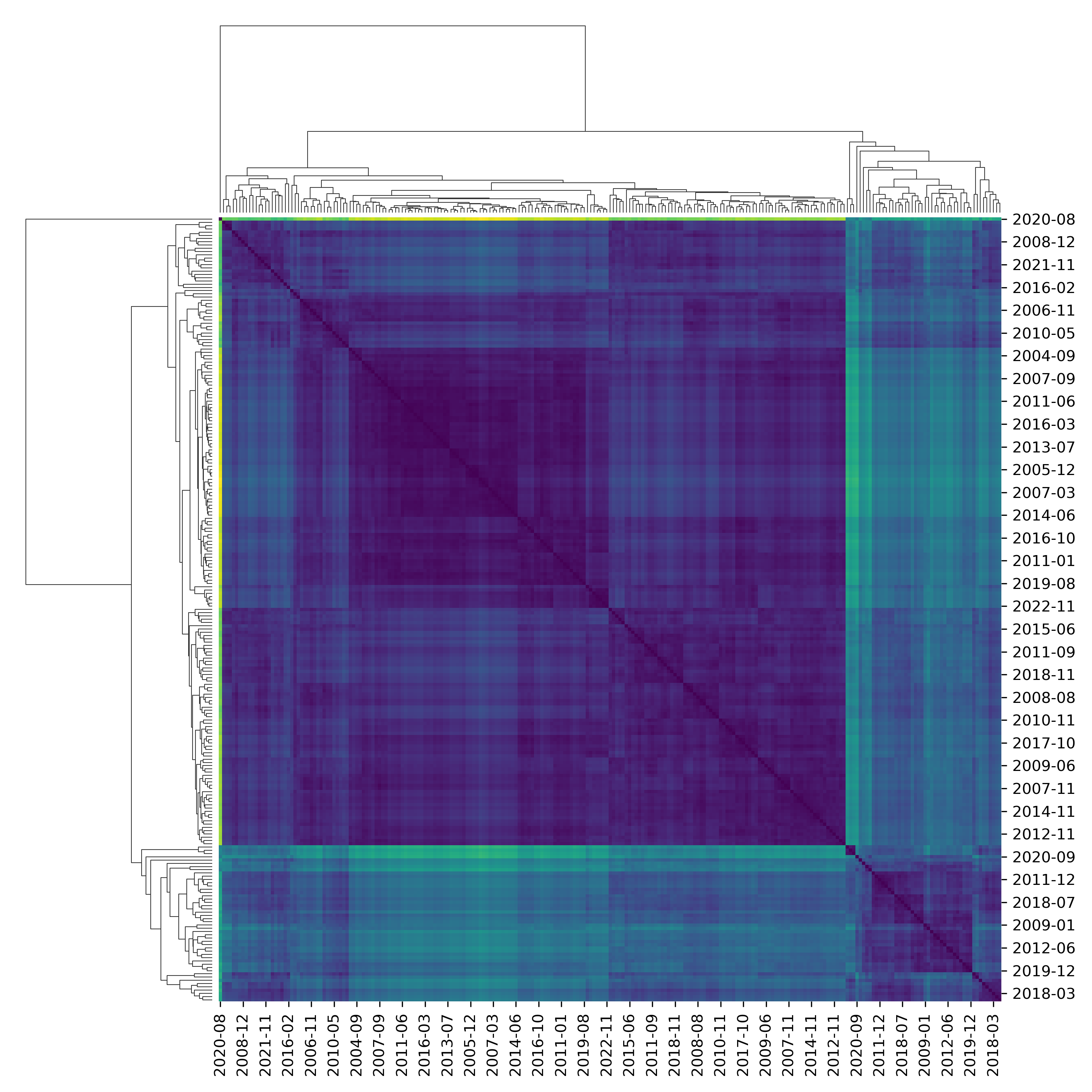}
\caption{Hierarchical clustering on $W(H_{t_1}, H_{t_2})$, defined in (\ref{eq:Wass_portfolio}), measuring the discrepancy between portfolio-weighted distributions of normalized market capitalizations in different months. This is effectively a weighted version of Figure \ref{fig:D_ST_Clustering}.}
\label{fig:dst_portfolio_weights}
\end{figure}

First, we analyze Figures \ref{fig:Heatmap_weight} and \ref{fig:weight_clustering} dedicated to representing the portfolio weights and relationships between them. Figure \ref{fig:Heatmap_weight} displays the time-varying vector of optimal weight coefficients for the market portfolio. This is an interesting object to study, as we can decouple the two opposing forces at work in the portfolio's objective function: decreasing the denominator of the Sharpe ratio by spreading risk across a variety of uncorrelated stocks, and the tendency for stocks with inherent affinity to perform well at specific instances in time. For example, when technology or mining stocks are rallying, typically passive investor money buoys the entire sector with indiscriminate capital inflows. Visual inspection of the figure suggests that a large portion of equities have sparse weight vectors over the analysis window, with several communities of stocks exhibiting spikes in their optimal weighting at coincident periods.

To determine the inherent similarity in evolutionary optimal weight vectors, we apply hierarchical clustering on the $L^1$ norm between these trajectories. This is shown in Figure \ref{fig:weight_clustering}. The outlier on the far left of the dendrogram is Netflix, which is attributed to a sub-cluster also consisting of other large technology companies such as Apple and Amazon. Given that these stocks would have similar levels of market beta, and co-exist in many passive products such as ETFs and thematic portfolios, it is unsurprising to learn that their weight vectors are highly similar. Surrounding the technology sub-cluster is a healthcare and biotechnology themed subcluster consisting of Regeneron pharmaceuticals, Vertex pharmaceutical and Intuitive Surgical. To the far right of the dendrogram we see Nvidia and Advanced Micro Devices clustering together too. It is quite clear that equities belonging to similar sectors and those exposed to the same broad macroeconomic themes (technology, biotech and pharmaceuticals, AI hardware, and so on) generate highly similar weight vectors. This would suggest that when forming Sharpe-optimal portfolios from the market, the association between stocks with high risk-adjusted returns is a more powerful component than spreading risk across equities with uncorrelated returns. This insight may have meaningful implications for investors wishing to optimize their portfolio in a dynamic way, while driving alpha consistently.

Next, we turn to the  Figures \ref{fig:portfolio_weighted_size}, \ref{fig:portfolio_normalized_weights}, and \ref{fig:portfolio_Gini}, which depict the various time-varying linear and nonlinear functionals of optimal portfolios we have defined. In Figure \ref{fig:portfolio_weighted_size}, we display the market cap weighted exposure of optimal portfolios $\bar{\nu}(t)$ in blue against the market average $\mu(t)$ in orange. Alternatively, the blue curve $\bar{\nu}(t)$ is the market cap exposure of a portfolio with an optimized Sharpe ratio, whereas the orange curve is the market cap of an equally weighted portfolio. In particular, when $\bar{\nu}(t)$ exceeds $\mu(t)$, the portfolio is more exposed to large companies than average; investors could interpret this as a measure of performance for a "company size" factor. Historically, there had always been an association between "growth" and small-cap stocks and "value" and large-cap stocks. Thus, a large value of $\bar{\nu}(t)$ may have been synonymous with a need for including higher quality stocks within an optimal investor portfolio. However, the emergence of market darling technology companies such as Nvidia, AMD and others has resulted in many larger companies being less appealing on traditional valuation metrics such as price-to-earnings, price-to-book and enterprise value-to-EBITDA. We see visually in our figure that for the majority of the analysis window, the optimal portfolio is underexposed with respect to market capitalization, suggesting a general preference on average for smaller companies. This echoes the historical finding of \cite{Reinganum1999}. This theme reverts closer to the present day, where the optimal portfolio is usually overexposed to larger companies. This is unsurprising given the massive stock price rally in large-cap technology and AI companies.

We now turn to Figure \ref{fig:portfolio_normalized_weights} to more closely study this over- and under-exposure phenomenon by graphing the ratio $\frac{\bar{\nu}(t)}{\mu(t)}$.  There are several notable takeaways in the figure. First, we see that Figure \ref{fig:portfolio_normalized_weights} is less than 1 for the majority of the period studied. There are three notable periods of exception: the global financial crisis (GFC), COVID-19 and the present day. During all three of these periods, optimal portfolios should include a larger proportion of larger companies. We remark that the reasons for these exceptional periods are quite distinct. For the GFC and COVID-19, larger companies typically had lower beta and protected portfolios from significant drawdowns. However, the most recent period in the market is in stark contrast to this, where  $\frac{\bar{\nu}(t)}{\mu(t)}$ trends up primarily due to the strong equity market performance of large-cap companies and the perceived scale benefit thereof, particularly in the technology sector.

These observations are interestingly complemented in Figure \ref{fig:portfolio_Gini}, which displays the time-varying nonlinear functional of optimal portfolio Gini $g(t)$ in blue, along with the Gini coefficient of the entire market (with respect to market cap) in orange. The latter is the same curve $G(t)$ previously graphed in Figure \ref{fig:Gini}. First, we can see the portfolio Gini exhibits far more variability than the market average Gini. This reveals that the level of market cap inequity in an optimal portfolio can change markedly over time. In fact, it is frequently close to "maximal inequality", where all the market cap is captured in a single asset with all the other values being close to zero.

Second, we reveal a complementary trend to that of the previous paragraph: the portfolio Gini coefficient is, broadly speaking, very high for the first half of the time window (and certainly higher than the market Gini), but then is rather low in the second half, closer to the present day. The general trajectory of this curve, which trends downward over time, has some interesting implications. The optimal portfolio, viewed as a distribution of market capitalization values, exhibits greater inequality earlier on, and greater equality closer to the present. This may be counterintuitive, but mathematically, it is likely because more recent portfolios are more concentrated at the top (cf Figures \ref{fig:Concentration_ratios} and \ref{fig:portfolio_weighted_size}  and) and this may actually reduce the Gini coefficient relative to the start of the time window, where stocks of all sorts of market cap sizes feature in the portfolio. This has meaningful implications for investors seeking to outperform the index. First, when this measure is high, it is important that investors (and their investment policy statements) are not constrained to diversify their portfolios with respect to size. It is clear that company size (like other investment themes such as growth, value, quality, and so on) is a factor with a tendency to trend. The second implication for investors is that when the portfolio Gini is declining significantly (indicating a more equitable distribution of weights across stocks), investors may be more likely to generate alpha via bottom-up stock selection rather than asset allocation strategies. That is, if we have a uniform distribution over stock performance with respect to company market caps, investors may wish to identify other discriminatory axes where the underlying phenomena has a more pronounced distributional difference. For instance, in some sectors with bifurcated outcomes, obvious winners and losers may emerge and "stock-picking" strategies may be more beneficial in driving portfolio returns.

In Figure \ref{fig:dst_portfolio_weights}, we display the hierarchical clustering on the Wasserstein metric  between market cap values of optimal portfolios weighted by portfolio weights. This figure measures the similarity between all points in time with respect to the distribution of market capitalization of companies that appear in Sharpe-optimally constructed portfolios. The cluster structure of the dendrogram has meaningful implications regarding the evolutionary regimes of optimal portfolios with respect to company size. A dendrogram with high similarity between all points in time would suggest that there may be one dominant regime, or sub-regimes with minimal variation to adjacent periods, and a persistent portfolio composition (with respect to company size) may suffice. Alternatively, a complex and intricate cluster structure would indicate that investors seeking to continually outperform should update their portfolio allocation regularly, and change the portfolio's size composition significantly upon each reallocation step. Of course, it is important to note that for a regime identification (and portfolio reallocation) strategy to be successful, the temporal constituents within each cluster must be fairly compact. That is, they must be sampled from a tight support on the temporal distribution. 

Figure \ref{fig:dst_portfolio_weights} displays one dominant cluster, with two fairly pronounced subclusters. There is also a smaller secondary cluster at the bottom right of the figure. Thus, one could surmise that there are at least two distinct regimes in terms of size-based optimal portfolio construction. However, given the wide variability of dates within each compact cluster and subcluster, this would indicate limited success in a dynamic trading strategy based on switching portfolio allocations based on a determined market \textit{market state} or \textit{market regime}.

\section{Conclusion}
\label{sec:Conclusion}

This paper has analyzed the structure of the entire market and several aspects of optimal portfolios (chosen by Sharpe ratio), each primarily with respect to market capitalization and over time. Through various original avenues of analysis, several coherent themes have emerged. Most clearly, notable changes in market structure and optimal portfolios have emerged since the start of 2020. As seen in Figure \ref{fig:Concentration_ratios}, the market cap concentration of the top stocks increased notably since the start of 2020, most of all among the top $k=5$. The Gini coefficient of the market reversed its downward trend over 2004-2020 and increased once more (Figure \ref{fig:Gini}). The distributions of normalized market cap values changed significantly in Figure \ref{fig:D_ST_Clustering}, with a sharp break in cluster structure pre- and post-2020. Changes were also visible in several aspects of optimal portfolios. Before 2020, optimal portfolios generally exhibited underexposure to market capitalization (Figures \ref{fig:portfolio_weighted_size} and \ref{fig:portfolio_normalized_weights}), reflecting historical trends where small-cap equities yielded stronger returns most of the time \cite{Reinganum1999}. However, this switched after 2020, since which optimal portfolios have more commonly exhibited overexposure to large market capitalization. At the same time, the Gini coefficient of optimal portfolios has consistently decreased, which could be, counterintuitively, caused by greater concentration of portfolios at the top of the market, rather than being more distributed throughout the market.

Aside from the above primary finding, there are several meaningful implications for investors in this paper. First, retail investors may acquire most market exposure inexpensively (with far fewer than 100 stocks). Unfortunately, this may then place such investors at a greater risk of excessive concentration, holding a disproportionate amount of their wealth in just a few companies. This could be especially concerning given that this small number of companies may be associated with the same sector, such as technology and AI. Interestingly, as the investment business continues to become more commoditized with lower marginal costs and barriers to entry for retail investors entering the market with increasing levels of sophistication, this could drive a self-enforcing phenomenon where size asymmetry may be exacerbated.

Second, our Sharpe ratio portfolio optimization analysis studies the strength of the two counteracting forces in the dynamic Sharpe ratio objective function (diversification vs investing in stocks with strong returns momentum). We observe that investing in equities with higher risk-adjusted returns is a stronger force than diversifying across equities with uncorrelated returns. Practically, this has meant that for retail investors seeking to generate portfolio returns on par with institutional investors - simply buying large stocks that are trending up has proved to be a successful (if not near-optimal) solution, despite the aforementioned risks in doing so. Third, our analysis of optimal portfolio Gini coefficient reveals that for dynamic investors, investment policy statements must be highly flexible and allow investors to overexpose their portfolios to specific businesses with significant pricing momentum.

In Section \ref{sec:portfolio_functionals}, we have introduced a theory of functionals (linear and nonlinear) of optimal portfolios, and direct our study as to how this relates to distribution of market capitalization. This fits naturally within the broader work of quantitative finance and time series analysis, combining the study of time-varying optimization in various market regimes with the added consideration of company size. Although the methods introduced in this paper explicitly consider market capitalization in the context of portfolio construction, these approaches can be generalized to consider other critical factors for investors to track over time. Future research could do the reverse: incorporate other factors such as market capitalization into optimal portfolio selection and track evolving aspects of these portfolios chosen in alternative ways. Our clustering methodology in Section \ref{sec:portfolio_functionals} could also be extended into a more mathematical approach for testing the temporal consistency of clusters and thus determine regimes of optimal trading strategies.

\bibliographystyle{_elsarticle-num-names}
\bibliography{__NEWREFS}
\biboptions{sort&compress}

\end{document}